\documentclass[]{aa}

\usepackage{rotating}
\usepackage{graphicx}
\usepackage{sidecap}
\usepackage{subcaption}
\usepackage{url}
\usepackage[colorlinks=true]{hyperref}
\usepackage{soul}
\hypersetup{allcolors=blue}
\usepackage{microtype} 

\usepackage{lipsum}
\usepackage{txfonts}
\usepackage[]{xcolor}
\usepackage{float}

\bibpunct{(}{)}{;}{a}{}{,}
\DeclareUnicodeCharacter{00A0}{~}
\usepackage[us,12hr]{datetime}
\usepackage{ulem}

\newcommand{\zbf}{\mathbf{z}}
\newcommand{\xbf}{\mathbf{x}}
\newcommand{\thetabf}{\boldsymbol{\theta}}
\newcommand{\fbf}{\mathbf{f}}

\newcommand{\kms}{\,km\,s$^{-1}$}

\newcommand{\cair} {\ion{Ca}{ii}~8542\,\AA\:}

\usepackage{cancel}
\usepackage{color}

\definecolor{deepmagenta}{rgb}{0.8, 0.0, 0.8}
\definecolor{green}{rgb}{0.0, 0.5, 0.1}

\begin{document}

    \title{Bayesian Stokes inversion with Normalizing flows}

   \author{C. J. D\'iaz Baso
          \inst{1}
          \and
          A. Asensio Ramos
          \inst{2,3}
          \and
          J. de la Cruz Rodr\'iguez
          \inst{1}
          }

   \institute{Institute for Solar Physics, Dept. of Astronomy, Stockholm University, AlbaNova University Centre, SE-10691 Stockholm, Sweden \email{carlos.diaz@astro.su.se}
   \and Instituto de Astrof\'isica de Canarias, C/V\'{\i}a L\'actea s/n, E-38205 La Laguna, Tenerife, Spain
   \and Departamento de Astrof\'isica, Universidad de La Laguna, E-38206 La Laguna, Tenerife, Spain\
             }

   \date{Draft: compiled on \today\ at \currenttime~UT}

   \authorrunning{D\'iaz Baso et al.}

\abstract
{

Stokes inversion techniques are very powerful methods for obtaining information on the thermodynamic and magnetic properties of solar and stellar atmospheres. In recent years, very sophisticated inversion codes have been developed that are now routinely applied to spectro-polarimetric observations. Most of these inversion codes are designed for finding an optimum solution to the nonlinear inverse problem. However, to obtain the location of potentially multimodal cases (ambiguities), the degeneracies, and the uncertainties of each parameter inferred from the inversions, algorithms such as Markov chain Monte Carlo (MCMC), require to evaluate the likelihood of the model thousand of times and are computationally costly. Variational methods are a quick alternative to Monte Carlo methods by approximating the posterior distribution by a parametrized distribution. In this study, we introduce a highly flexible variational inference method to perform fast Bayesian inference, known as normalizing flows. Normalizing flows are a set of invertible, differentiable, and parametric transformations that convert a simple distribution into an approximation of any other complex distribution. If the transformations are conditioned on observations, the normalizing flows can be trained to return Bayesian posterior probability estimates for any observation. We illustrate the ability of the method using a simple Milne-Eddington model and a complex non-LTE inversion. The method is extremely general and other more complex forward models can be applied. The training procedure need only be performed once for a given prior parameter space and the resulting network can then generate samples describing the posterior distribution several orders of magnitude faster than existing techniques.

}



   \maketitle


\section{Introduction}\label{sec:intro}

Through the analysis of spectra and their polarization, we have been able to infer the properties of the solar and stellar atmospheres. To infer the stratification of physical properties as a function of depth, we compare the emergent spectra given by a solar model with observations. By modifying the physical parameters that define the model atmosphere, we can find a specific configuration that resembles the observations and obtain a physical interpretation of the origin of the observed phenomena. This process is commonly known as spectropolarimetric inversion and nowadays is routinely used in solar physics to extract physical information from spectropolarimetric observations. 

At present, there are different methods for obtaining the most probable atmospheric structure responsible for producing the observed spectra (see \citealt{delToro2016} and \citealt{Jaime2017} for extensive reviews). The traditional way for finding the optimum solution is the use of a gradient search minimization algorithm, normally of second order, such as the Levenberg-Marquardt method (\citealt{Levenberg1944, Marquardt1963}). There are a collection of techniques that use the gradient of the merit function to drive the solution in the direction of the minimum by reducing the difference between the forward calculated spectrum and the observed one. They usually require only a few forward calculations of the synthetic spectra to converge, but the global minimum can only be guaranteed if the merit function is convex.

To have complete knowledge of the parameter space (the location of the global minimum if it exists, whether there are degeneracies or multiple solutions that can equally reproduce the observations, and to have a proper estimation of the uncertainty in the solution), the Bayesian framework has to be used. The posterior distribution of the model parameters conditioned on the observations encodes all the relevant information of the inference. Computing this high-dimensional posterior distribution turns out to be complex and one has to rely on efficient stochastic sampling techniques such as Markov Chain Monte Carlo (MCMC; \citealt{Metropolis1953}) and nested sampling \citep{Skilling2004}.

These sampling methods have been one of the pillars of Bayesian analysis in solar physics \citep{Asensio2007, Arregui2018} not only to sample complicated posterior distributions for parameter inference but also to compute marginal posterior for model comparison. Even though they are compelling and new algorithms have been developed to obtain the overall shape of the parameter space when multiple solutions exist \citep{DiazBaso2019}, they require many forward calculations and are therefore computationally very costly, which limits their applicability even in relatively simple atmospheric models.

Currently, although successful, the use of gradient-based methods has found an obstacle when applied to two-dimensional fields of view with millions of pixels. Inverting such large maps often requires the use of supercomputers running parallelized inversion codes for many hours \citep[e.g.,][]{Dalda2019}. This is especially relevant for inversions in non-local thermodynamic equilibrium (NLTE), for which the forward problem is computationally heavy. The use of artificial neural networks (ANN) to learn the non-linear mapping between the observed Stokes parameters and the stratification of solar physical parameters has shown a large improvement in speed and robustness to noise compared to classic gradient search inversion codes \citep{Carroll2001, Socas2005, Asensio2019, Socas2021}. Given that the inverse problem is often not bijective (there are very similar spectra that emerge from very different physical parameters), traditional neural networks struggle in cases where multiple solutions exist. A solution to this is to use neural networks as emulators which mimic the forward modeling and accelerate standard MCMC posterior sampling methods in Bayesian inversions \citep[e.g.,][]{BayesClumpy2009}.

A very promising alternative method for Bayesian inference is variational inference, where the true distribution of the solution is approximated with a simpler analytical distribution \citep{Asensio2017, DiazBaso2019noise}, more convenient to work with. By assuming a distribution instead of a single solution for the physical parameters, we can capture the uncertainties and, for instance, whether several solutions are compatible with the observations. To improve this approximation, mixture models are usually used as a composition of many components of the same or different families. However, they require the evaluation of each component, their optimization is not always stable, and they only work well in cases where the posterior is simple.

In this study, we leverage a variational inference method known as normalizing flows  \citep{tabak2013, Rippel2013, Dinh2014, Rezende2015} to do Bayesian inference for the physical atmospheric parameters from spectropolarimetric data. Normalizing flows are a set of parametrized transformations that can convert a simple and analytically known distribution (for instance, a standard normal distribution) into a more complex distribution. Normalizing flows use neural networks to represent this complex relation. When these normalizing flows are conditioned on the observations, they can approximate the posterior distribution for every observation very efficiently. They also provide all the tools to rapidly sample from the posterior and compute the ensuing probability of every sample.

Given their recent conception and development, there are only a few examples of astrophysical applications, such as estimating continua spectra of quasars \citep{Reiman2020}, constraining distance estimates of nearby stars \citep{Cranmer2019} or inferring physical properties of black holes using gravitational waves \citep{Green2020}. In solar physics, \cite{Osborne2019} showed the use of invertible neural networks, a particular modification of normalizing flows where the inverse and forward models are learned at the same time. The information that is lost during the forward model (which makes the inverse model ill-defined) is re-injected again by using an ad-hoc latent vector. The forward and inverse models are made consistent by imposing a cycle consistency. In our experience, although it is a promising direction, the cycle consistency makes both models somewhat hard to train. In this study, we prefer to approximate both models separately. The inverse problem is treated probabilistically with a normalizing flow, while the forward model is approximated with a standard neural network. We have found this approach to be stable and fast to train.

Therefore, we propose to apply normalizing flows to spectro-polarimetric observations to perform Bayesian inference faster than the classical gradient search algorithms with all the added information of the parameter space given by the posterior distribution. We present an automated inference framework based on neural density estimation, where the fundamental task is to estimate a posterior distribution from pre-computed samples from a physical forward model. The paper is organized as follows. We start with a brief introduction to normalizing flows (Sec.~\ref{sec:nflow}), their basic principles, and how we implement the new approach with solar observations. Later we show the application of the normalizing flow for Bayesian inference on two examples with different complexity (Sec.~\ref{sec:ME} \& \ref{sec:NLTE}) and verify the accuracy of this approach (Sec.~\ref{sec:accuracy} \& \ref{sec:fov}). Finally, we provide a brief discussion about the implications of this work and outline potential extensions and improvements (Sec.~\ref{sec:conclusions}).

\section{Normalizing flows}\label{sec:nflow}

\begin{figure*}
\centering
\includegraphics[width=\linewidth]{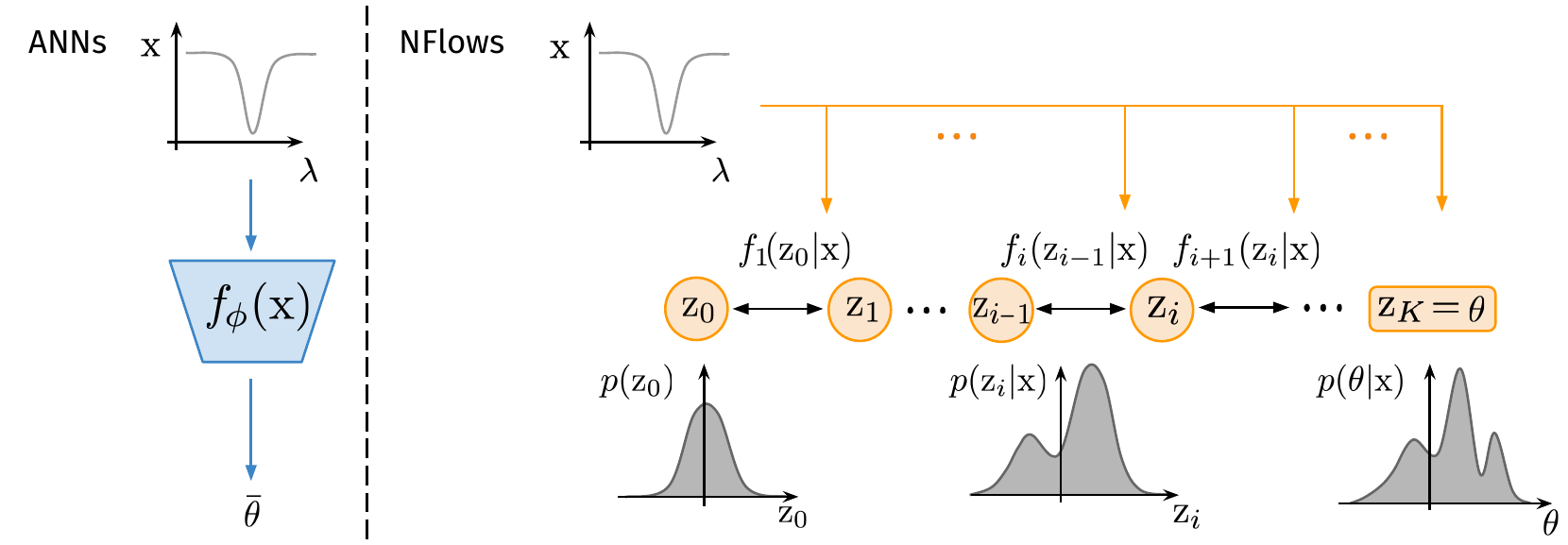} 
\caption{Comparison between an artificial neural network and a conditional normalizing flow with parameters $\phi$. The output of the classical artificial neural network is an average value over all the solutions $\overline{\theta}$, however the normalizing flow transforms a simple distribution conditioned on the data to obtain the full probability distribution of the target parameter $\theta$.}
\label{fig:sketch}
\end{figure*}

\subsection{Amortized Variational Inference}

Bayesian inference relies on calculating the posterior probability distribution function $p(\thetabf|\xbf)$ of a set of $M$ physical model parameters $\thetabf$ that are used to describe a given observation $\xbf$ with $N$ data points. This quantity can be obtained by direct application of the Bayes theorem \citep{Bayes63}:
\begin{equation}
    p(\thetabf|\xbf) = \frac{p(\xbf|\thetabf) p(\thetabf)}{p(\xbf)},
\end{equation}
where $p(\xbf|\thetabf)$ is the likelihood, which measures the probability that the data $\xbf$ were obtained (measured) assuming given values for the model parameters $\thetabf$, while $p(\thetabf)$ is the prior distribution of all possible model parameters. The quantity $p(\xbf)$ is the so-called evidence or marginal posterior, which normalizes the posterior probability distribution. 

As mentioned, the calculation of the posterior can be complex and one must rely on efficient stochastic sampling techniques, such as MCMCs, which require many model evaluations. On the other hand, variational inference is a faster alternative that tries to approximate the posterior with a simpler parametrized distribution, translating the inference problem into an optimization problem. However, both methods work on a single instance of the problem. This means that when we want to perform inference on a large number of observations, we have to run the same method several times which, at the end, can be significant in terms of computing time. In this way, we are not exploiting the global knowledge about the problem and evaluating many times the same model with different data.

Recent advances in deep learning have promoted the development of new algorithms based on amortized neural posterior estimation \citep[ANPE;][]{Kyle2019}, i.e., we can find a function that maps the observed data to the variational parameters of our approximation. Once we have this function, performing inference on a new observation is as simple as evaluating this function and plugging the output into our variational approximation. This function must be flexible enough to learn the global representation of the data and neural networks are good at learning features from the data directly. Thus, neural networks can be optimized to approximate the posterior distribution $p(\thetabf|\xbf)$ by conditioning the resulting distribution on observations that have been precomputed over our prior parameter space $p(\thetabf)$. After training the density estimator, we can evaluate the posterior of new observations efficiently. The approach presented in this paper belongs to the ANPE framework.

\subsection{Normalizing flows}

Normalizing flows approximate the posterior distribution ${p}(\thetabf|\xbf)$ by transforming a simple probability distribution $p_z(\zbf)$ into a complex one by applying an invertible and differentiable transformation $\thetabf = \mathbf{f}(\mathbf{z})$. In practice, we can construct a flow-based model by implementing $\mathbf{f}=\mathbf{f_\phi}$ with a neural network with parameters $\phi$ and take the base distribution $p_z(\zbf)$ to be simple, typically a multivariate standard normal distribution. This base distribution will act as a latent space of hidden independent variables. If the observed variables are also independent, the transformation will be just a scaling of each of them. On the contrary, if the joint distributions of the observed variables show high complexity (correlations, multiple maxima, etc), the latent variables will be mixed to reproduce such distributions.
More precisely, the resulting probability distribution is computed by applying the change of variables formula from probability theory \citep{Rudin2006}:
\begin{equation}\label{eq:chng_vrbl}
    p_\phi(\thetabf|\xbf)=p_z(\mathbf{z})\left|\det \dfrac{\partial \mathbf{f_\phi}}{\partial \mathbf{z}}\right|^{-1}
    =p_z(\mathbf{f_\phi^{-1}}(\thetabf))\left|\det \dfrac{\partial \mathbf{f_\phi}^{-1}}{\partial \thetabf}\right|
\end{equation}
where the first factor represents the probability density for the base distribution ($p_z$) evaluated at $\mathbf{f_\phi^{-1}}(\thetabf)$, while the second factor is the absolute value of the Jacobian determinant and accounts for the change in the volume due to the transformation. This factor forces the total integrated probability of the new distribution to be unity. The transformation $\mathbf{f_\phi}$ expands, contracts, deforms and shifts the probability space to morph the initial (base) distribution into the target.

A flexible transformation can be obtained by composing several simple transformations, which can produce a very complex distributions. An important property of invertible and differentiable transformations is that if we compose $K$ transforms {$\mathbf{f}=(\mathbf{f}_{1}\circ \mathbf{f}_{2}\circ\cdots\circ \mathbf{f}_{K})$}, their inverse can also be decomposed in the components {$\mathbf{f^{-1}}=(\mathbf{f}^{-1}_{K}\circ\cdots\circ \mathbf{f}^{-1}_{2}\circ \mathbf{f}^{-1}_{1})$} and the Jacobian determinant is the product of the determinant of each component. Therefore the log-probability of the overall transformation is then:
\begin{equation}
     \log p_\phi(\thetabf|\xbf) \simeq \log p(\mathbf{z}_K|\xbf) = \log p_z(\mathbf{z}_0) + \sum^K_{k=1}\log\left|\det\frac{\partial \fbf^{-1}_k}{\partial {\mathbf{z}_{k-1}}}\right|
     \label{eq:normalizing_flow}
\end{equation}
where $\mathbf{z}_k=\mathbf{f}_k(\mathbf{z}_{k-1})$. The term normalizing flow is intimately related to the compositional character described above. The term "flow" refers to the trajectory that a collection of samples follow as they are gradually transformed. The term "normalizing" refers to the inverse trajectory that transforms a collection of samples from a complex distribution and makes them converge towards a prescribed distribution, often taken to be the normal distribution.

Figure~\ref{fig:sketch} illustrates the difference between the classical neural-assisted Stokes inversion methods \citep[e.g.][]{Asensio2019} (upper panel) and the probabilistic approach that normalizing flows offer (lower panel). In the classical case, a neural network produces a point estimate of the parameters $\overline{\theta}$ from the observed Stokes parameters. This point estimate is not really representative of any of the potential solutions in degenerate cases. In the probabilistic case, a normalizing flow transforms a standard-normal base distribution together with the observations into a complex target posterior density through several simple transformations. This final distribution properly captures all solutions compatible with the observations, even if they are multimodal.

Normalizing flows are trained by minimizing the discrepancy (or divergence) between the posterior distribution and the variational approximation of Eq.~(\ref{eq:normalizing_flow}). The most popular choice is the Kullback-Leibler divergence \citep{KL1951}. Assuming that our dataset has $D$ samples $\{\boldsymbol{\theta_i},\mathbf{x_i}\}$ with physical parameters drawn from the prior $p(\boldsymbol{\theta})$ and observations generated from our physics-based forward model $p(\mathbf{x}|\boldsymbol{\theta})$, minimizing the Kullback-Leibler divergence is equivalent to minimizing the negative log-likelihood:
\begin{equation}
    \mathcal{L_\phi} = -\frac{1}{D}\sum^D_{i=1} \log p_\phi(\boldsymbol{\theta_i}|\mathbf{x_i})
\end{equation}
The minimization is carried out using gradient-based optimization methods by calculating the gradient of $\mathcal{L_\phi}$ with respect to the parameters of the normalizing flows, $\phi$. During the training, the normalizing flow will be optimized to map the samples from the unknown distributions to a standard normal distribution. If that is performed successfully, the inverse transformation enables us to sample  the posterior by simply extracting values from the normal distribution and applying the learned inverse transformations.

From this result, we can already appreciate the power of normalizing flows on density estimation: we do not need to evaluate the likelihood of our data $p(\xbf|\thetabf)$ (or the model) during training, we only need to provide samples. This is also the goal of simulation-based inference \citep[SBI,][]{Kyle2019}, to perform Bayesian inference when the evaluation of the likelihood is not possible because of mathematical or computational reasons.

\subsection{Invertible transformations}

A normalizing flow should satisfy several conditions to be of practical application. It should be: 1) invertible, 2) expressive enough to model any desired distribution, and 3) computationally efficient for calculating both forward and inverse transforms and the associated Jacobian determinants. Among the different families of transformations $\mathbf{f_\phi}$, we use a transformation known as coupling neural splines flows \citep{Dinh2014, Muller2018, Durkan2019} which have been demonstrated to be effective at representing complex densities, are fast to train, and fast to evaluate (see \citealt{Papamakarios2019Review} and \citealt{Kobyzev2019Review} for extensive reviews).

The idea behind the coupling transform was introduced by \citet{Dinh2014} and consists of dividing the input variable (of dimension $Q$) into two parts and applying an invertible transformation $\mathbf{g}$ to the second half ($\mathbf{z}_{q+1:Q}$), whose parameters are a function of the first half (i.e., $\mathbf{z}_{1:q}$). Such transformations have a lower triangular Jacobian whose determinant is just the product of the diagonal elements, allowing to create faster normalizing flows. The output vector $\mathbf{o}$ of a coupling flow is given by:
\begin{align}\label{eq:cplng_lyr}
\mathbf{o}_{1:q} &= \mathbf{z}_{1:q} \nonumber \\
\mathbf{o}_{q+1:Q} &= \mathbf{g}_{({\mathbf{z}_{1:q}})}(\mathbf{z}_{q+1:Q}),
\end{align}
where $\mathbf{g}_{({\mathbf{z}_{1:q}})}$ is an invertible, element-wise transformation whose internal parameters have been computed based on $\mathbf{z}_{1:q}$ and in our case (conditionals flows) also on the observed data $\mathbf{x}$. The final output of the transformation is then $\mathbf{o}=[\mathbf{o}_{1:q}, \mathbf{o}_{q+1:Q}]$. Since coupling layers leave unmodified $\mathbf{z}_{1:q}$, one needs to shuffle the order of the input in each step by using a permutation layer, so these two halves do not remain independent throughout the network.

For the coupling transformation $\mathbf{g}$, we have chosen a family of very expressive functions based on monotonically increasing splines \citep{Muller2018, Durkan2019}. They have shown large flexibility when modeling multi-modal or quasi-discontinuous densities. A spline is a piece-wise function that is specified by the value at some key points called knots. The location, value, and derivative of the spline at the knots for each dimension in $\mathbf{o}_{q+1:Q}$ are calculated with a neural network. Since each resulting distribution (and therefore each transformation) will depend on the observed data, the neural network will have the input $[\mathbf{z}_{1:q},\mathbf{x}]$.

\begin{figure*}
\centering
\includegraphics[width=0.91\linewidth]{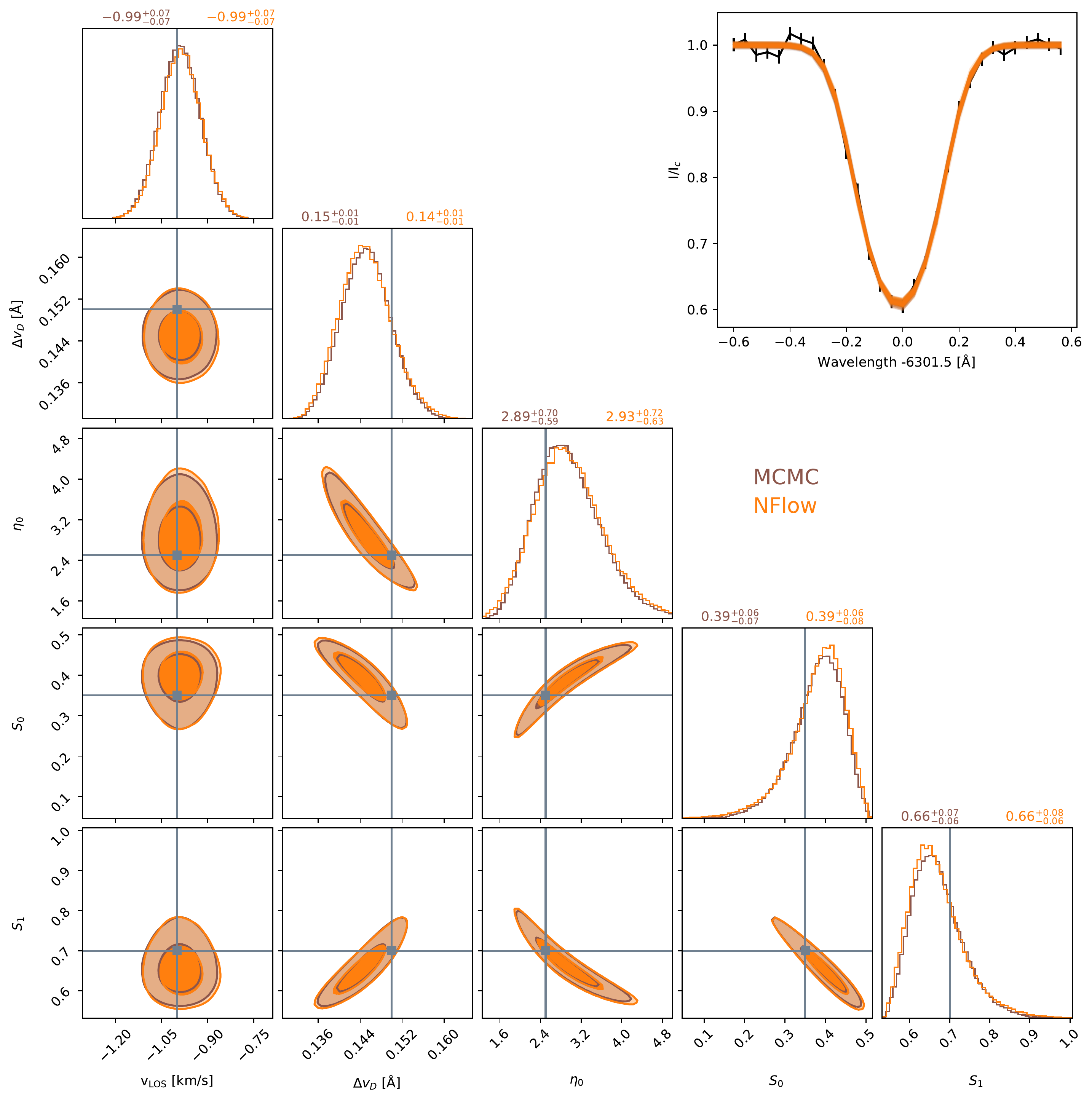}
\caption{Joint (below the diagonal) and marginal (in the diagonal) posterior distributions for the physical parameters involved in the Milne-Eddington model. The label on top of each column provides the median and the uncertainty defined by the percentiles 16 and 84 (equivalent to the standard 1$\sigma$ uncertainty in the Gaussian case). Also, the contours are shown at 1 and 2 sigmas. The original values of the parameters are indicated with
gray dots and vertical/horizontal lines.} \label{fig:ME_corner}
\end{figure*}

\begin{figure*}
\centering
\includegraphics[width=0.49\linewidth]{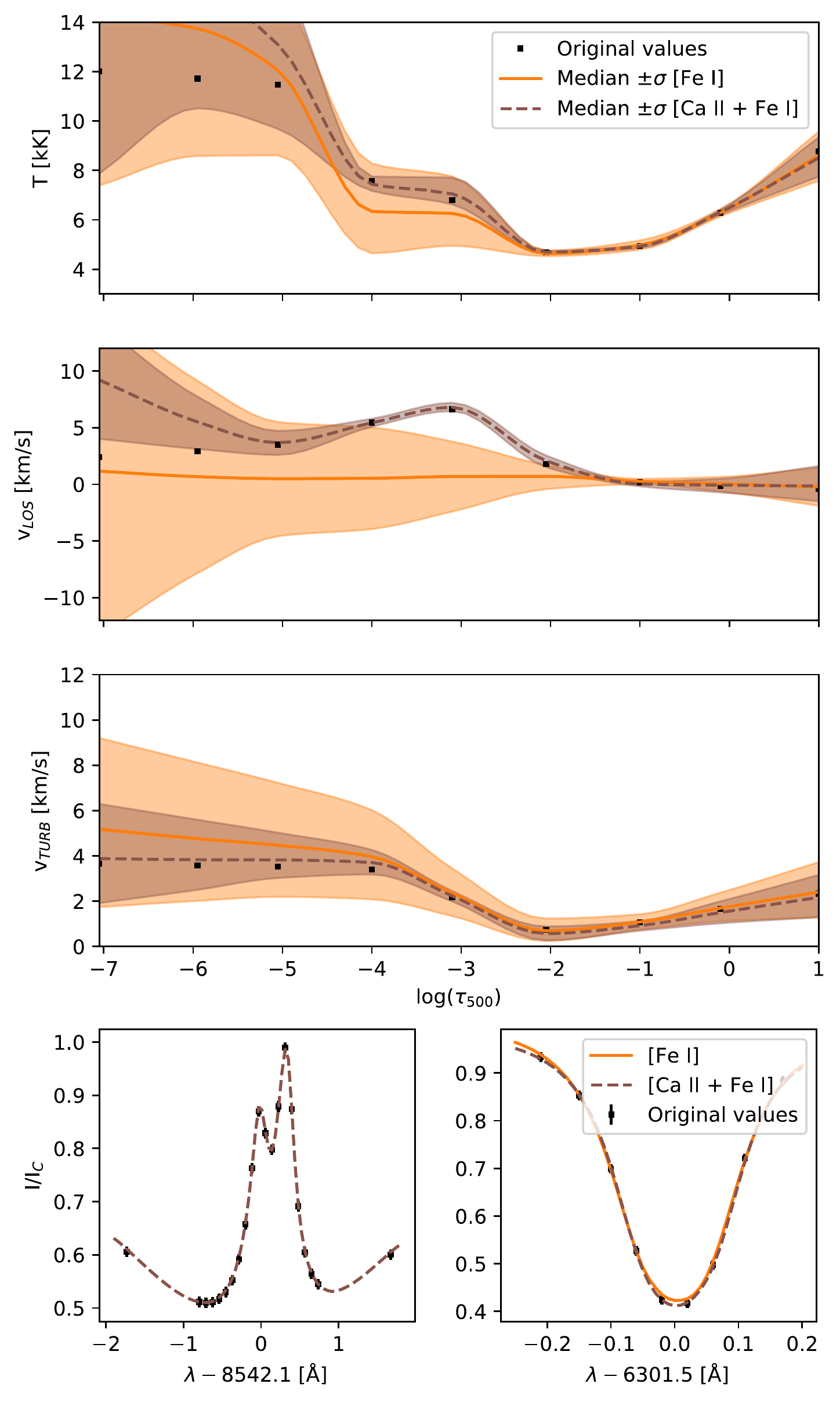}
\includegraphics[width=0.49\linewidth]{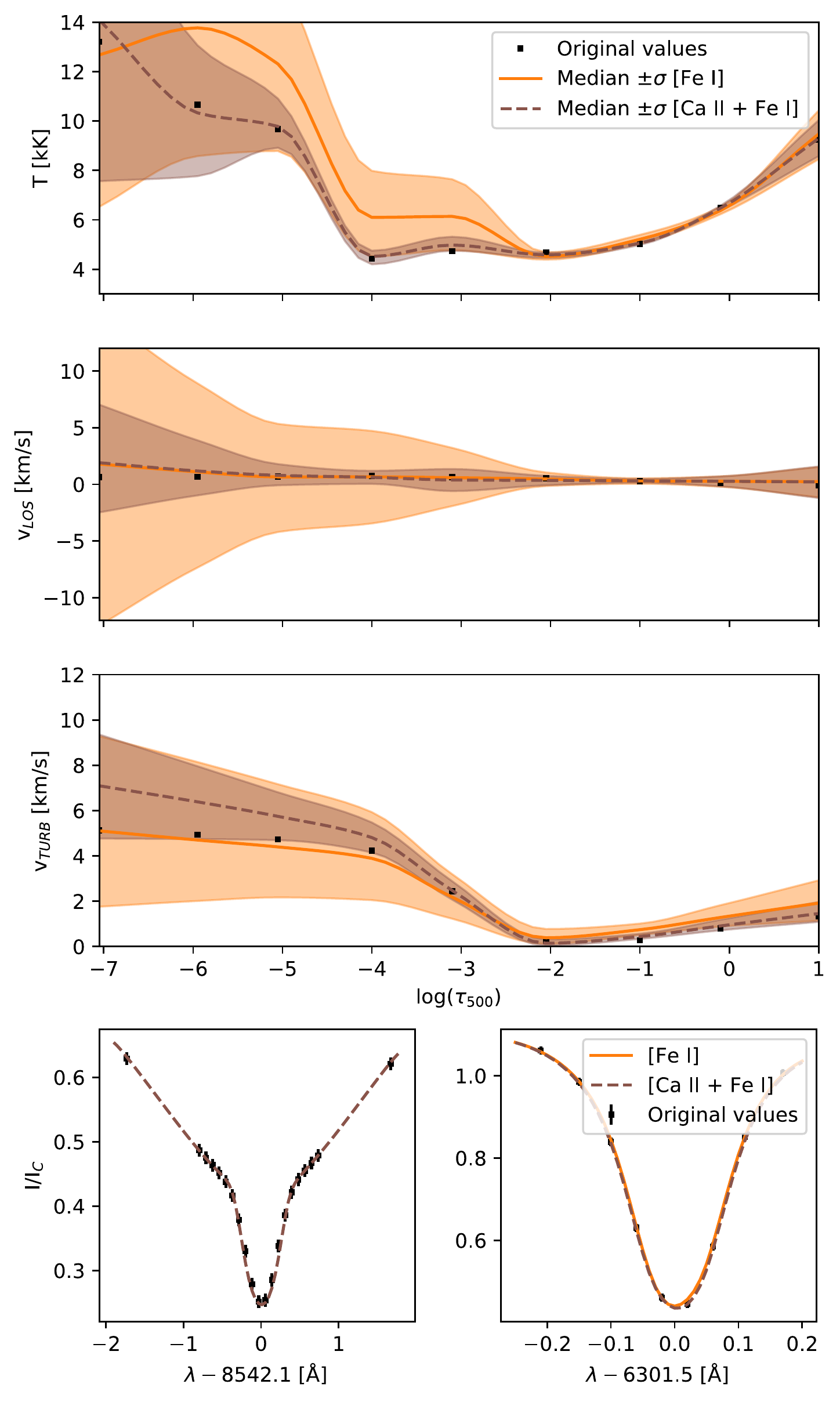}
\caption{Atmospheric stratification inferred by the normalizing flow for two examples. In each column, the orange solution is inferred only using the \ion{Fe}{i} line while the brown solution also uses the \ion{Ca}{ii} profile. The lowest row shows the original intensity values together with the synthetic calculation from the maximum a-posteriori solution.}
\label{fig:nlte}
\end{figure*}

\subsection{Network architecture and training details}

We use the implementation of normalizing flows in \texttt{PyTorch} \citep{PyTorch} available in \texttt{nflows} \citep{nflows} which allows for a wide variety of transformations to be used, among them the neural spline flows for coupling transforms. As suggested by \cite{Green2020}, we include invertible linear transforms together with a permutation layer before each coupling transform allowing all of the parameters to interact with each other. In summary, we have used an architecture that is a concatenation of blocks of an invertible linear transformation using the LU-decomposition \citep{Kingma2018} with a rational-quadratic spline transform \citep[RQ,][]{Durkan2019} where a residual network \citep[ResNet,][]{ResNet2015} is used to calculate the parameters of the splines. Our \texttt{PyTorch} implementation can be found in the following repository:  \url{https://github.com/cdiazbas/bayesflows}. We have also included a basic example to illustrate the method.

For the first simple case, a flow with 5 coupling transformations, 5 residual blocks, and 32 neurons per layer was enough. In the second case, we need at least 15 coupling layers, 10 residual blocks, and 64 neurons per layer. We trained the models for {500} epochs with a batch size of {100}. We have used a learning rate of {$10^{-4}$} and the Adam optimizer \citep{Kingma2014}. During training, we reserved 10\% of our training set for validation. An augmentation scheme based on applying different realizations of Gaussian noise to the data is applied during training, thus increasing the effective size of the training set.

\section{Simple case: Milne-Eddington atmosphere}\label{sec:ME}

As a first example, we show the capabilities of the normalizing flows in a case where the forward modeling is fast enough to allow comparison with the exact solution obtained with an MCMC method. For this case, we choose the Milne-Eddington \citep{Auer1977} solution of the radiative transfer problem as a baseline. Focusing only on Stokes $I$, $\thetabf$ is five-dimensional and contains the physical parameters of relevance that describe the intensity profile of a spectral line: the macroscopic velocity $v\rm _{LOS}$, the Doppler width $\Delta v\rm _D$, the line-to-continuum opacity ratio $\eta_0$, and the two parameters of the source function $S_0$ and $S_1$.

The normalizing flow is trained using an appropriate training set. To this end, we generate $10^6$ training pairs ($\theta_i,x_i$) by drawing $\theta_i$ using a uniform prior for all the variables in the following ranges: ${v\rm _{LOS}} =[-3.0,3.0]$ \kms, $\Delta v\rm _D =[0.05,0.2]$ \AA, ${\rm \eta_0} =[0.0,5.0]$, $S_0 =[0.0,1.0]$, $S_1 =[0.0,1.0]$. We have simulated the photospheric \ion{Fe}{i} 6301.5\AA\ line using the Milne-Eddington model\footnote{\url{https://github.com/aasensio/milne}} and assuming Gaussian noise of {$\sigma=8\cdot10^{-3}$} in continuum units. Once the normalizing flow is trained, we can carry out Bayesian inference for arbitrary observations. In order to demonstrate the speed of the inference, we point out that the normalizing flow produces samples of the posterior at a rate of {20000} per second. In the following we show the result for a synthetic profile with input parameters {$v_{\rm LOS}=-1.0$ \kms, $\Delta v_D = 0.15$ \AA, $\eta_0=2.5$, $S_0=0.35$, and $S_1=0.7$}. This profile was chosen on purpose to highlight the flexibility of the neural network when working with complex distributions and strong degeneracies. To verify the accuracy of the posterior inference we compare the result of the normalizing flow against a Markov Chain Monte Carlo computed with the \texttt{emcee} sampler \citep{emcee2013} using a Gaussian likelihood.

The comparison of the posterior distributions generated by the neural network (light orange) and \texttt{emcee} (dark orange) is shown in Fig.~\ref{fig:ME_corner}. The normalizing flow is doing a very good job at approximating the posterior, with both distributions clearly in very close agreement. For strongly degenerate parameters, such as ${\Delta v\rm _D}$ and ${\rm \eta_0}$, we recover a typical joint banana-shaped posterior. For highly correlated parameters, we find ridge-shaped distributions, like between the parameters ${\rm S_0}$ and ${\rm S_1}$. Given these strong degeneracies, a classical inversion method based on artificial neural networks will not perform correctly as there are multiple solutions for the same profile. The upper right panel of the same figure shows the synthetic spectra using samples from the posterior distribution of both methods. This test is known as the posterior predictive check and helps us understand whether the model is appropriate. It is clear from these results that the predictive profiles of each color are almost indistinguishable. They also lie within the uncertainty produced by the noise on the intensity profile we have evaluated.

\begin{figure*}
\centering
\includegraphics[width=0.49\linewidth]{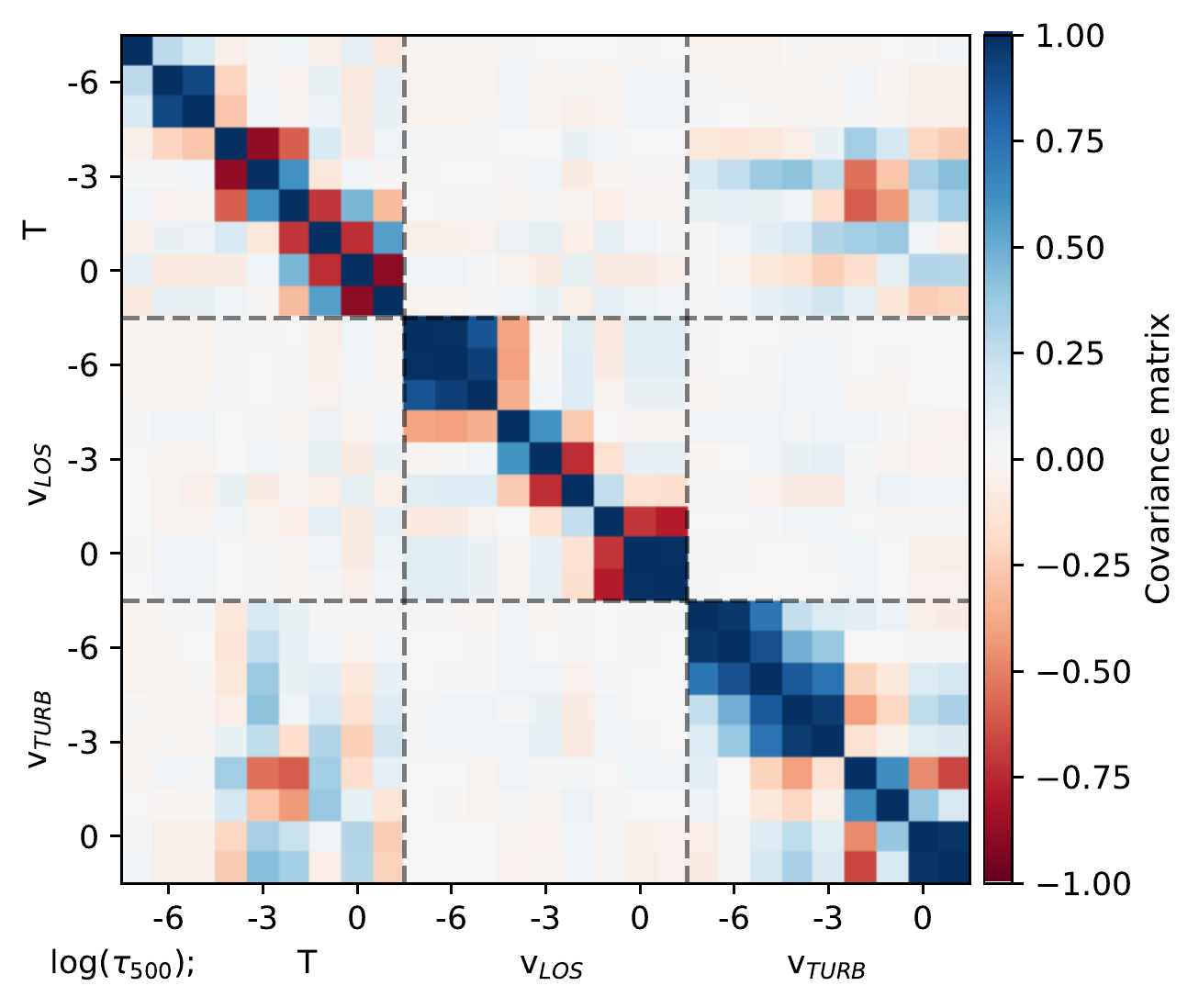}
\includegraphics[width=0.49\linewidth]{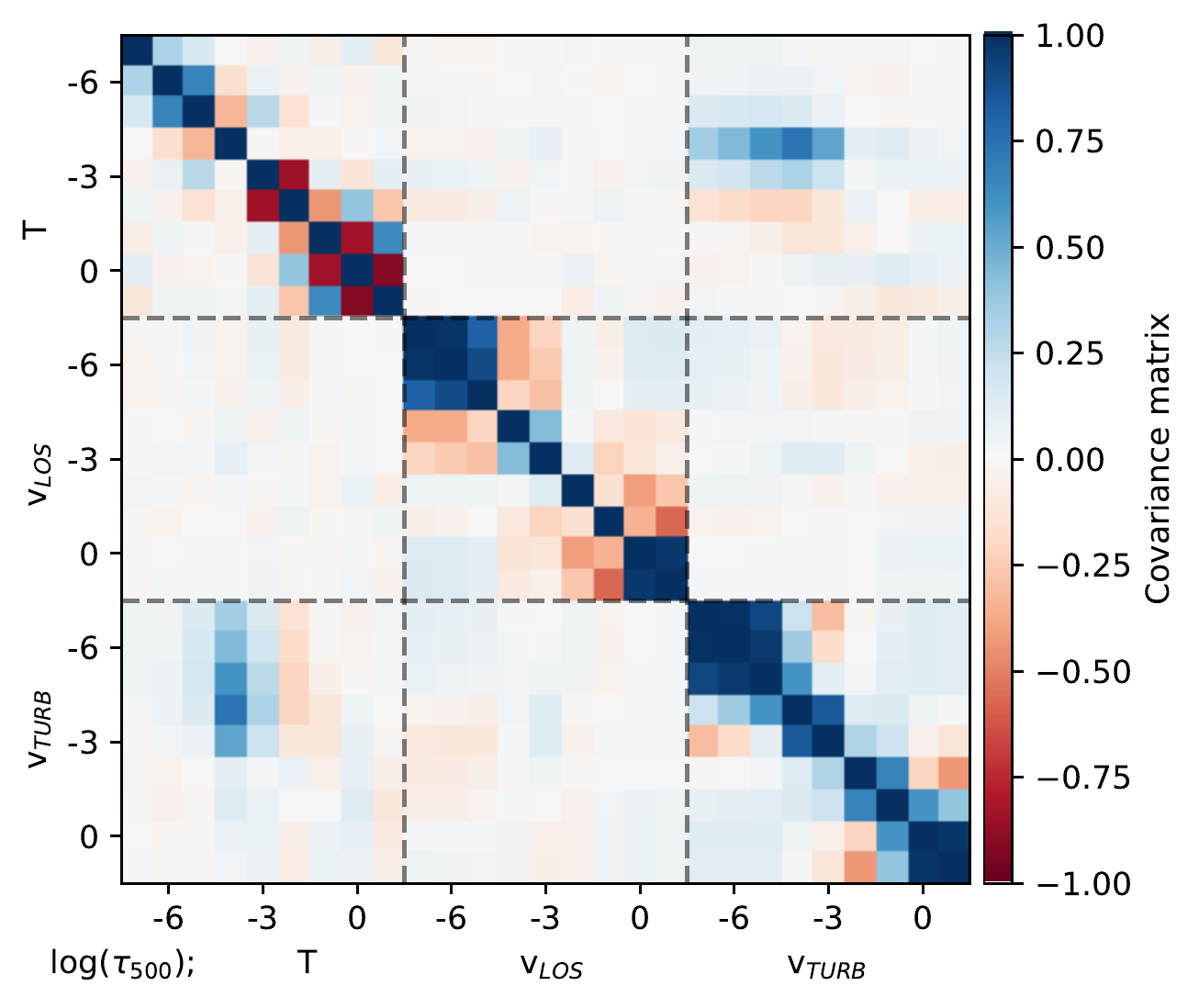}
\caption{Correlation matrices calculated for the inferred atmospheric stratification for the emission (left)
and absorption (right) profiles. Blue/red indicates positive/negative correlations, respectively. The optical depth increases towards the right and downwards, so each pair of physical quantities has the top of the atmosphere in the upper left corner and the bottom in the lower right corner.} 
\label{fig:cov}
\end{figure*}

\begin{figure}
\centering
\includegraphics[width=1.0\linewidth]{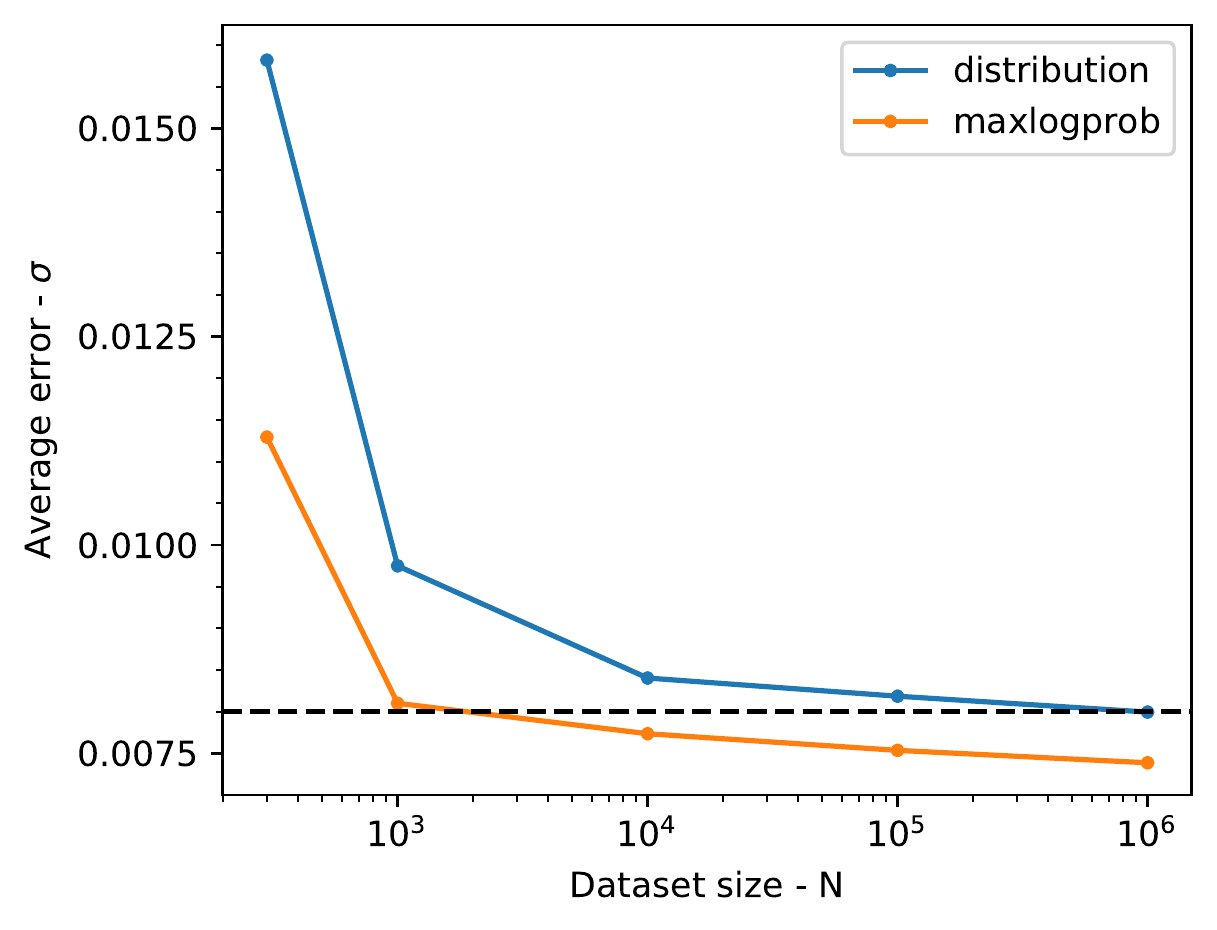}
\includegraphics[width=1.0\linewidth]{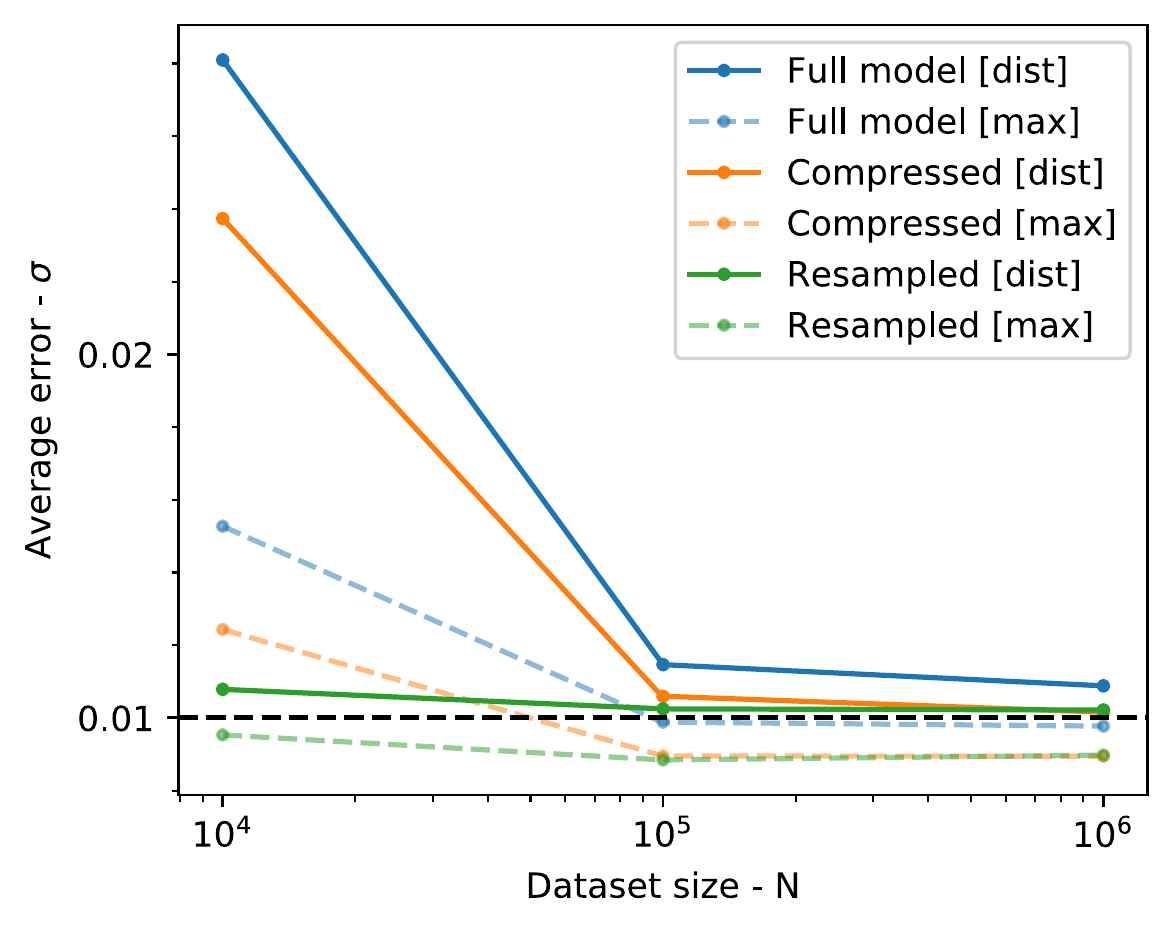}
\caption{Upper panel: statistical properties of the posterior predictive check for the ME case.
The blue curve averages over the full posterior distribution, while the orange curve
shows the difference with respect to the mode of the posterior. Lower panel: the 
same results for the NLTE case. We also display the effect of compressing the model
with an autoencoder and applying the resampling strategy.} 
\label{fig:error}
\end{figure}

\begin{figure*}
\centering
\includegraphics[height=0.96\textheight]{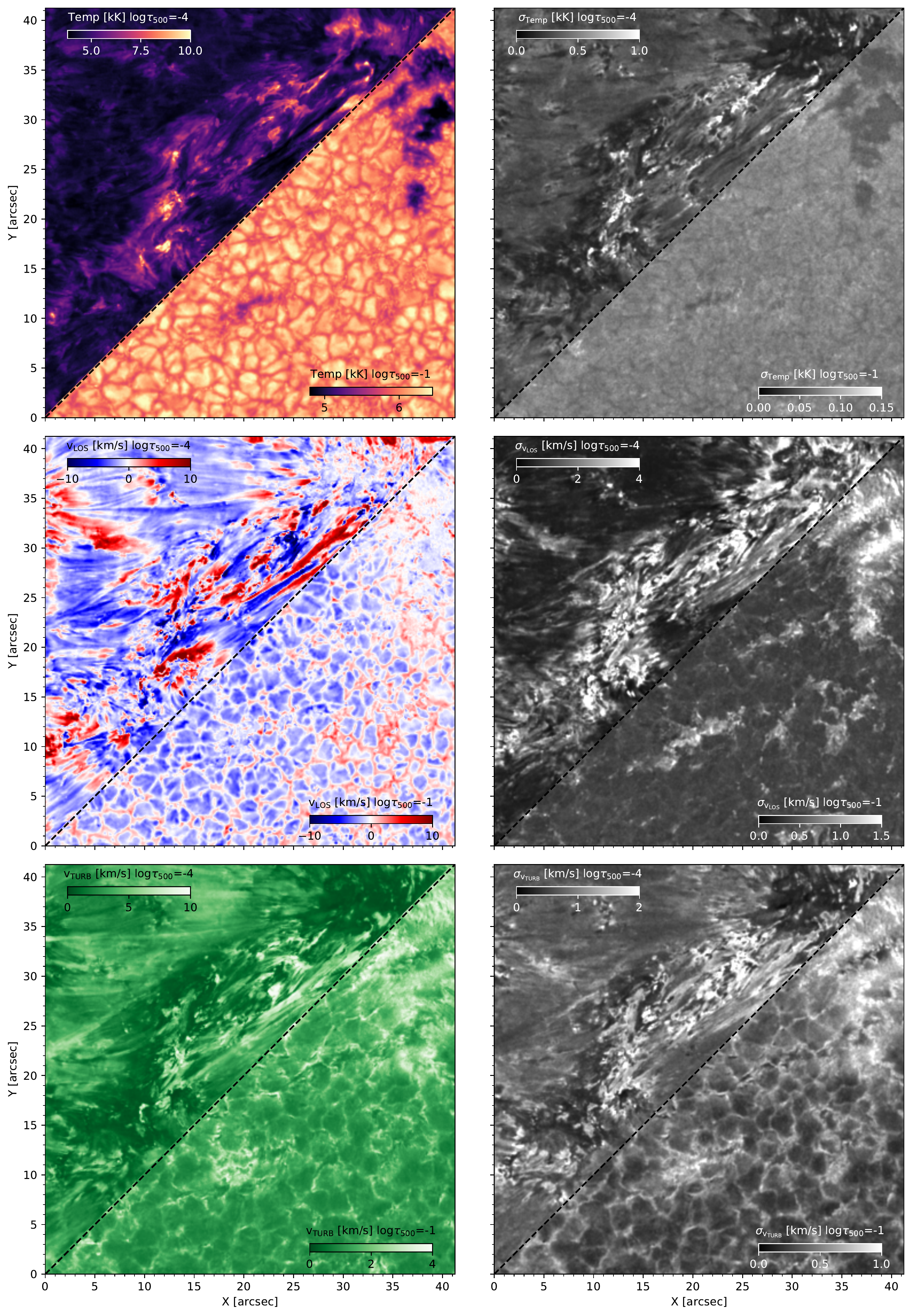}
\caption{Atmospheric structure of the FOV as inferred from the inversion. The left column shows the temperature, the LOS velocity, and the microturbulent velocity at two layers for half of the FOV. The right column shows the associated uncertainty for the same quantities and layers.} 
\label{fig:fov}
\end{figure*}

\begin{figure}
\centering
\includegraphics[width=\linewidth]{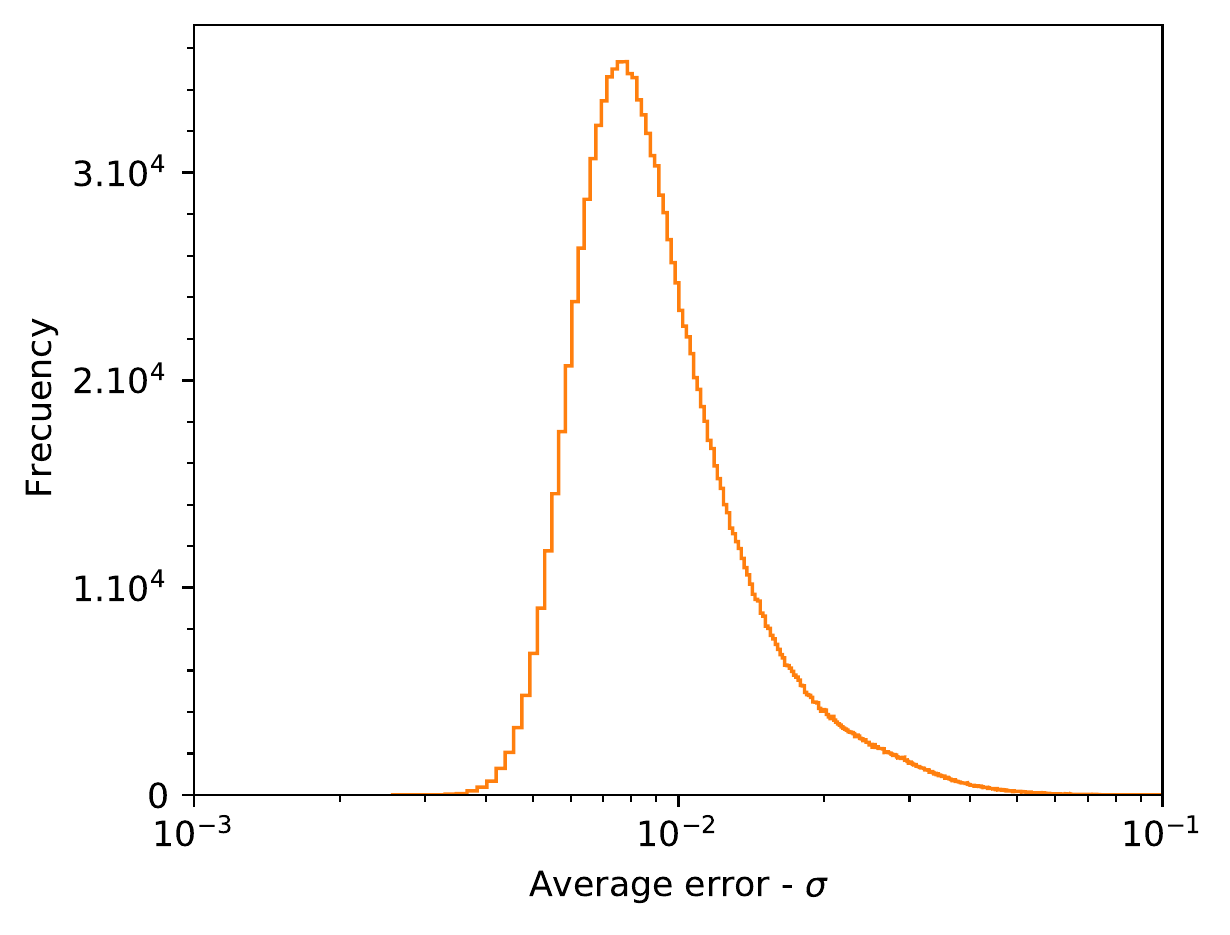}
\caption{Distribution of the error in intensity between the synthetic and the observed profiles for the inversion of the FOV in Fig.~\ref{fig:fov}.} 
\label{fig:chi2}
\end{figure}

\section{Complex case: NLTE with stratification}\label{sec:NLTE}
Computing the posterior distribution in complex and realistic models turns out to be particularly difficult, for several reasons. First, complex models often require lots of computing power, so the number of samples of any MCMC method will be limited by the available computing time. An example of this is the formation of chromospheric spectral lines under NLTE conditions, which requires the joint solution of the radiative transfer and statistical equilibrium equations for the atoms under consideration. Second, complex models also require more model parameters (such as the whole atmospheric stratification with height), which increases the dimensionality of the posterior distribution. Finally, complex problems often contain non-linear calculations that can lead to very cumbersome posterior distributions, with strong degeneracies and ambiguities. In the case of the NLTE problem, the information encoded in the posterior strongly depends on the specific observed spectral line. Some heights in the atmosphere are constrained by the observations, while other heights remain partially or completely unconstrained. This produces a posterior distribution with very different widths for different parameters of the model.

As a showcase for our approach, in this second example, we have simulated a case in which we have simultaneously observed the photospheric \ion{Fe}{i} 6301.5\,\AA\ line and the chromospheric \ion{Ca}{ii} 8542\,\AA\ line. This configuration is commonly used because it facilitates studying events occurring both in the photosphere and the lower chromosphere \citep{2019ApJ...870...88E,2019A&A...627A.101V, Kianfar2020,Diaz2021, Yadav2021}. It is also widespread because it is one of the most common instrumental configurations of the CRISP instrument on the Swedish 1-m Solar Telescope \citep[SST,][]{Scharmer2003, Scharmer2008}.

The normalizing flow model needs to be trained for this specific problem. To this end, and to create a diverse set of samples of solar-like stratifications and intensity profiles, we have started from the solar stratifications inferred in \cite{Diaz2021}. These models were inferred from spectropolarimetric observations under some model assumptions. Therefore, the posterior distributions produced by our model will be dependent on this specific a-priori information (more details below). Anyway, we consider that the assumption of a prior is beneficial since it will act as a guide to avoid non-realistic stratifications. The model can always be trained with a different prior extracted, for instance, from state-of-the-art numerical simulations.

The stratifications of the training set were extracted at nine equidistant locations from $\log(\tau_{500})=-7.0$ to $\log(\tau_{500})=+1.0$, where $\tau_{500}$ is the optical depth measured at 500 nm. These are the locations at which we infer the posterior distributions. We computed the mean and covariance matrix of the temperature, velocity, and microturbulent velocity obtained from the results of \cite{Diaz2021} at these locations. We created $10^6$ new stratifications by sampling from multivariate normal distributions with the computed means and covariances. Each physical quantity is drawn independently. To increase the diversity of the dataset, we have increased the standard deviation at each $\log(\tau_{500})$ location by a factor 2-6. The samples are then interpolated to 54 depth points using Bezier splines and encapsulated in a model with zero magnetic field. The gas pressure at the upper boundary is assumed to be P$_{\rm top}$= 1 dyn cm$^{-2}$. The density and gas pressure stratifications are computed by assuming hydrostatic equilibrium (HE). The spectra were synthesized using the multi-atom, multi-line, NLTE inversion code STiC \citep{delaCruz2016,delaCruz2019_STiC}. This code is built on top of an optimized version of the RH code \citep{Uitenbroek2001} to solve the atom population densities assuming statistical equilibrium and plane-parallel geometry. The radiative transport equation is solved using cubic Bezier solvers \citep{delaCruz2013} and includes an equation of state extracted from the SME code \citep{Piskunov2017}. We treated the \ion{Ca}{ii} atom in NLTE with the \cair line in complete frequency redistribution (CRD). The \ion{Fe}{i} 6301.5\AA\ line was also modeled in NLTE, hence accounting for a complex atmosphere that could affect its formation. We have then degraded each spectral line to the spectral resolution of the CRISP instrument at the SST telescope and an uncorrelated Gaussian noise with a standard deviation of 10$^{-2}$ in units of the continuum intensity was added.

We train two different models to capture the improvement in the inference of the stratification when more spectral lines are included. The first model only uses observations of the \ion{Fe}{i} line, while the second model uses both the \ion{Fe}{i} and \ion{Ca}{ii} lines together. The trained models are used to infer the stratification of physical properties for two different examples: one with strong emission in the chromospheric line produced by an increase in the temperature in this region and another example with the core of both lines in absorption. 

The results of the inference are shown in Fig.~\ref{fig:nlte}. The left panels refer to the case in emission, while the case in absorption is displayed in the right panels. The top three panels show the stratification with optical depth of the temperature, line-of-sight velocity, and turbulent velocity. Black squares indicate the original stratification used to synthesize the line profiles. The solid orange and dashed brown lines show the median value estimated from the posterior distribution when considering only the photospheric line or both lines, respectively. The shaded regions mark the corresponding 68\% confidence interval. As expected, the inference that considers both lines can recover with high accuracy the whole stratification, whereas using only the \ion{Fe}{i} line yields a model where only the photosphere is recovered, with a large uncertainty towards the upper atmosphere. This result shows that the normalizing flow is able to learn the range of sensitivity of each spectral line just by looking at the examples of the database. This sensitivity is model-dependent and a direct consequence is the presence of a larger uncertainty in the upper atmosphere when the profile is in emission with respect to that found when the profile is in absorption. This difference in the width of the distribution comes from the fact that large temperatures ionize the \ion{Ca}{ii} atoms and the line becomes less sensitive to the local physical conditions in the upper layers where the temperature is large \citep{Diaz2021}. Although the \ion{Fe}{i} line contains very limited information about the upper chromosphere, the inference outputs an increasing temperature and not a completely uncertain posterior. The reason for this is that the posterior is simply recovering the prior we have used, which has larger temperatures towards the upper atmosphere. As in any Bayesian inference, the prior gives an important constraint on how solar-like stratifications are expected, giving low probabilities to unrealistic values where the line contains no information.

Although uncertainties are easy to visualize in the marginal posterior distributions at each location, the joint distributions are difficult to visualize for a multidimensional model. Joint distributions give interesting information about how different parameters are correlated in the inference, clearly pointing out the presence of ambiguities. To obtain an approximate insight, the joint distribution can be summarized by showing the correlation matrix. These correlation matrices are shown in Fig.~\ref{fig:cov} for the two examples of Fig.~\ref{fig:nlte}. Non-zero correlation is found both inside the same physical parameter (intra-parameter) and between different parameters (inter-parameter). Part of the intra-parameter correlation is a consequence of the assumed prior, which forces smooth stratifications. The checkerboard pattern found in the temperature indicates that reductions in the temperature at some locations can be compensated by increases in other locations. This oscillatory behavior can appear during the classical inversion process if too many degrees of freedom are used, and regularization is usually used to avoid erratic behavior \citep{delaCruz2019_STiC,delaCruz2019_STiCB}. 

Arguably the most visible inter-parameter correlation happens between the temperature and the microturbulent velocity. This correlation is different in amplitude and sign between the two examples because it depends on the particular model stratification and the non-linear effect of the physical parameters in the spectra. In this case, the difference comes from the fact that an increase of temperature in the chromosphere when the line profile is in absorption will produce a narrower core and an increase of the microturbulence compensates for this difference (positive correlation). However, an increase in temperature in the emission profile will broaden the profile, and the microturbulence needs to be decreased (negative correlation). The location of this correlation also indicates that the absorption profile displays the response higher in the atmosphere when compared with the emission profile. In summary, the Bayesian inference with normalizing flows allows us to obtain a complete picture of the inferred stratification and the degeneracy between parameters.

\section{Validation and performance}\label{sec:accuracy}

Normalizing flows have the combined advantages of Bayesian inference and deep learning, by gaining access to the posterior distribution and being able to sample from it very fast.  However, it also inherits some of the limitations of the classical MCMC methods. For example, to obtain a sufficiently good representation of the posterior distribution with a standard MCMC method, it is required to densely sample the parameter space around the location of the mode, first to properly find it and then to estimate its shape. The samples from the training set have to be closer or similar to the expected width of the posterior distribution so that the normalizing flow is able to work properly. This is especially critical when the uncertainty in the observations is low because one expects very narrow posteriors. In this case, using a sparse training set leads to an overestimation of the posterior width.

We have quantified how the accuracy of our models depends on the size of the training set. To this end, we use the fact that, when the models extracted from the posterior distribution are used to re-synthesize the line profiles, they should be distributed according to the assumed sampling distribution. In our case, this sampling distribution is Gaussian with a standard deviation of 8$\cdot$10$^{-3}$ and 10$^{-2}$ in units of the continuum intensity for the ME and NLTE case, respectively. The upper panel of Fig.~\ref{fig:error} shows these results for the ME case. 

The blue curve has been calculated as follows: we have produced the corresponding posterior distributions of 1000 different spectra from the database, then for each posterior distribution we have extracted 1000 models and calculated the standard deviation between the original and the synthesized profiles, and finally we have calculated the average between all the standard deviation values.  This has been done for each normalizing flow that has been trained with a different training set, producing all blue curve points. The orange curve shows the same quantity but only for the maximum a-posteriori solution (evaluating only the solution with the highest probability).

The average error of the normalizing flow model decreases asymptotically with the size of the dataset towards the noise level of the target spectra (shown as a horizontal dashed black line). The speed at which this convergence takes place strongly depends on the complexity of the forward problem. In the Milne-Eddington example, the orange curve reaches the expected error already for a training set of 10$^3$ examples, although this number needs to be much higher when we check for convergence of the full posterior distribution. In the NLTE case, the model requires many more simulated examples to reach a similar error (see lower panel of Fig.~\ref{fig:error}). This behavior can also be witnessed during training: while the validation loss in the ME case saturates, the validation loss in the NLTE starts to worsen after {500} epochs, indicating that the number of samples is probably not large enough to avoid overfitting. This behavior is less noticeable with the largest training dataset indicating we are reaching the right number of samples. This overfitting disappears when a larger standard deviation is used in the Gaussian noise added to the profiles. In this case, the model is able to correctly reproduce the width of the posterior distribution with a smaller amount of training examples. We note that the size of the training set for a normalizing flow model only for \ion{Fe}{i} is similar to that of the ME case.

We have considered two possible procedures to reduce the effect of the size of the training set on the results of the model. The first procedure relies on using compression to reduce the dimensionality of the forward model. To this end, we use an autoencoder \citep{Hinton2006} to compress the 27 nodes (9 nodes for three physical variables) into a vector of dimension 20. After trial and error, we have found this value to be the minimum size of the bottleneck layer so that the error in the reconstruction of the stratifications is still smaller than the width of the posterior distributions. This autoencoder consists of an encoder, a bottleneck of dimension 20, and a decoder. Both the encoder and the decoder are fully connected ResNets with 5 residual blocks and 64 neurons per layer. The autoencoder is trained with the physical stratifications of the training set by minimizing the difference between the input and output of the autoencoder. Once trained, we use the encoder to produce compressed representations for the stratifications, which are used to train the normalizing flow. After training the normalizing flow, one can use the decoder of the autoencoder to produce physical stratifications from the samples of the flow. A more compact representation helps the normalizing flow to train faster, also performing better (see orange lines in Fig.~\ref{fig:error}). 

The second procedure reuses the samples from the normalizing flow and reweights them using importance sampling to produce a better approximation to the posterior distribution. This requires one to have access to the forward modeling, which is time-consuming in complex NLTE cases. One can also use a pre-trained neural network that works as an emulator of the forward modeling and produces much faster synthesis. In such case, the posterior distribution can be further improved by resampling according to the likelihood of the observations, producing the results of the green lines in Fig.~\ref{fig:error}. To this end, we re-weight our posterior samples so that the sampling distribution approaches the correct posterior. The importance sampling weights are, therefore:
\begin{equation}
    w = \frac{p(\thetabf|\xbf)}{ p_{\phi}(\thetabf|\xbf)} = \frac{p(\xbf|\thetabf) p(\thetabf)}{ p_{\phi}(\thetabf|\xbf)},
\end{equation}
where the likelihood is given by

\begin{equation}
    \log p(\xbf|\thetabf) = -\frac{1}{2} \sum_{j=1}^{N_\lambda} \left[ 
    \frac{\left( S(\lambda_j,\thetabf) - I(\lambda_j) \right)^2}{\sigma_{j}^2} - \frac{1}{2} \log 2 \pi \sigma_{j}^2 \right],
\end{equation}
where $\sigma_{j}^2$ is the wavelength-dependent noise variance, $S(\lambda,\thetabf)$ is the synthetic line profile for parameters $\thetabf$, while $I(\lambda)$ is the observed line profile. The number of wavelength points we consider is $N_\lambda$. We point out that this resampling process only works well when the posterior samples from the normalizing flow overlap with those of the real posterior distribution and the sampling of the real posterior is not very sparse. This is almost guaranteed in our case because our normalizing flow tends to overestimate the width of the posterior distribution but not for a large margin.

Given that the precision of the normalizing flow model increases with the size of the training set, one should consider this trade-off for a particular problem. For complex problems for which the difficulty of generating a large simulated training set is dominating, one should consider reducing the maximum achievable accuracy by using a larger standard deviation of the noise. This automatically produces broader posteriors. Anyway, we are confident that the normalizing flow model is an improvement over more classical MCMC methods. Any MCMC sampling method would require performing $10^{4}-10^{5}$ synthesis to correctly sample the posterior distribution for any new observation that needs to be analyzed. The amortized character of our model allows it to be applied seamlessly to new observations.

\section{Validation on large FOVs}\label{sec:fov}
We have also tested the trained normalizing flows on large fields of view. To qualitatively challenge this method on real data, we have chosen the observations analyzed in \citet{Leenaarts2018A&A}. The data was taken at a heliocentric angle $\mu$=1 and its wavelength sampling is the same used in the creation of the training set. This region was observed with the SST on 2016-09-19 at around 09:30 UT using the CRISP instrument. The field of view target is an active region with elongated granulation indicating ongoing flux emergence and enhanced brightness in the core of the chromospheric lines above the flux emergence (see  \citealt{Leenaarts2018A&A} for more details).

Assuming that our database contains a sufficiently diverse sample of observed profiles, we have applied the neural network to a field of view of approximately 42$\times$42 arcseconds (smaller than the original size for optimal visualization). The normalizing flow was applied pixel by pixel, and to speed up the inference only 50 samples have been obtained; enough to roughly estimate the width of the distribution. Figure~\ref{fig:fov} shows in the left and right columns the mean stratification and standard deviation of the samples at each pixel for each physical magnitude, respectively. The lower half of each panel shows the quantity at $\log(\tau_{500})=-1$ which shows the photosphere, and the upper half at $\log(\tau_{500})=-4$, which provides a view of the chromosphere. In the line-of-sight velocity maps, the blue color represents motions toward the observer while the red color represents motions away from the observer.

Overall the model performs well, generating coherent maps and reproducing the patterns previously found in gradient-based inversions \citep{Leenaarts2018A&A, Diaz2021}. We note that since our model does not include a magnetic field, all of the non-thermal broadening is reproduced with a higher value of the microturbulent velocity. The uncertainties tend to increase from the photosphere to the chromosphere, on average 80\,$\xrightarrow{}$500K in the temperature, 0.5\,$\xrightarrow{}$1.2\kms\ in the LOS velocity, and 0.4\,$\xrightarrow{}1.0$\kms\ in the microturbulent velocity. At a first glance at the two columns, there seems to be a clear correlation between the magnitudes and their uncertainty. This could be understood from the examples in the previous sections since emission profiles have higher chromospheric temperatures and therefore lower sensitivity in the upper layers. These locations also coincide with regions with high velocities and turbulent motions as a result of the interaction of the flux emergence with the pre-existing magnetic field. There are also exceptions such as the low photospheric velocity field in the umbra with higher uncertainty, probably caused by the difficulty to extract an accurate value from very wide profiles broadened due to the magnetic field. However, a more detailed analysis of each region is beyond the scope of this work.

To estimate the performance of the network on these data we re-synthesized the spectral lines from the mean stratification of each pixel. Figure~\ref{fig:chi2} shows a histogram of the average error of each pixel for the FOV. On average, the result is very good, with a peak value around $10^{-2}$, although with an extended tail in the distribution reaching higher values. These points with a higher error are associated with the most complex profiles in the interior of the region. This behavior is to be expected because although the training set has a large diversity of profiles, more complex profiles have stronger gradients in temperature and velocity and therefore would require a finer sampling of heights than the one used here.

We want to note that for this particular set we did not find significant differences among the synthetic spectra coming from the mean, median and maximum a-posteriori stratification (MAP). The reason could be the following: the posterior distribution is narrow where the spectral lines are very sensitive and all these values are very close, while for the regions of the atmosphere where the posterior is much wider and asymmetric (so MAP, median, and mean are different) the lines do not show a clear imprint  of the solar conditions at those locations. This result cannot be generalized to other multimodal cases such as the magnetic field solutions under the Zeeman or Hanle effects.

\section{Summary and conclusions}\label{sec:conclusions}

In this study, we have explored the usage of normalizing flows to accurately infer the posterior distribution of a solar model atmosphere (parameters, correlations, and uncertainties) from the interpretation of observed photospheric and chromospheric lines. Once the normalizing flow model is trained, the inference is extremely fast. We have also shown that the quality of the approximate posterior distribution depends on the size of the training set and that applying dimensionality reduction techniques makes the normalizing flow performs better. Rapid parameter estimation is critical if complex forward models are used to analyze a large amount of data that the next generation of telescopes such as the Daniel K. Inouye Solar Telescope \citep[DKIST;][]{DKIST2015} and the European Solar Telescope \citep[EST;][]{EST2016} will produce.

Compared with a classical inversion approach, that computes only single-point estimations of the solution, models based on normalizing flows have the ability to estimate uncertainties and multi-modal solutions. The latter is usually the case when a single spectral line is used and its shape can be explained by different combinations of the parameters. It is, therefore, crucial to use multi-line observations and characterize how they can constrain the inferred properties, as we have shown here.

There are still many ways in which we could improve the current implementation. First, a natural extension of this work would be to include the four Stokes parameters to infer the magnetic properties of our target of interest, while also setting more constraints in the rest of the physical parameters. Since ambiguous solutions often plague the inference of magnetic properties, the Bayesian framework is ideal for it. Second, our normalizing flow only works with a particular noise level, but it can be easily generalized to arbitrary uncertainties by, for instance, passing it as input in the conditioning of the normalizing flow \citep{2021arXiv210612594D}. An arguably better option would be to let the model infer its value \citep[e.g.,][]{DiazBaso2019}. Third, a more general model can be built by adding metadata such as the spectral resolution or the solar location as additional conditioning, alongside the observation. We are actively working on a version of the model that uses a summary network to condition the normalizing flow. This will enable the use of arbitrary wavelength grids \citep{2021arXiv210809266A}. All of the above together will allow us to create a general Bayesian inference tool for almost any arbitrary observation. 

Another problem we have shown is the difficulty in generating an appropriate training set. Recently, alternative training methods have been developed that sequentially propose samples from the training set to maximize model performance with a smaller number of samples (see \citealt{Lueckmann2021} for a comparison of different approaches), and that is an idea that we plan to explore in the future. Finally, the ability of normalizing flows to model probability distributions makes this method versatile and other applications such as anomaly detection or learning complex priors for more traditional Bayesian sampling methods \citep{Alsing2021} are also being considered.

\begin{acknowledgements}

{We would like to thank the anonymous referee for their comments and suggestions.}
CJDB thanks Adur Pastor Yabar, Flavio Calvo and Roberta Morosin for their comments.

AAR acknowledges financial support from the Spanish Ministerio de Ciencia, Innovaci\'on y Universidades through project PGC2018-102108-B-I00 and FEDER funds.
This project has received funding from the European Research Council (ERC) under the European Union's Horizon 2020 research and innovation program (SUNMAG, grant agreement 759548).
%
The Swedish 1-m Solar Telescope is operated on the island of La Palma by the Institute for Solar Physics of Stockholm University in the Spanish Observatorio del Roque de los Muchachos of the Instituto de Astrof\'isica de Canarias. The Institute for Solar Physics is supported by a grant for research infrastructures of national importance from the Swedish Research Council (registration number 2017-00625).
We acknowledge the community effort devoted to the development of the following open-source packages that were used in this work: numpy (\url{numpy.org}), matplotlib (\url{matplotlib.org}), scipy (\url{scipy.org}), astropy (\url{astropy.org}) and sunpy (\url{sunpy.org}).
%
%
This research has made use of NASA's Astrophysics Data System Bibliographic Services.
\end{acknowledgements}

\bibliographystyle{aa}

\end{document}